\pdfoutput=1
\documentclass[preprint,12pt]{elsarticle}

\usepackage{amssymb}
\usepackage{physics}
\usepackage[mathscr]{euscript}%
\usepackage[margin=2cm]{geometry}

\journal{ArXiv}

\begin{document}

\begin{frontmatter}



\title{Model-Free Forecasting of Partially Observable Spatiotemporally Chaotic Systems}


\author{Vikrant Gupta$^a$, Larry K. B. Li$^b$, Shiyi Chen$^a$, Minping Wan$^a$\\
{\small{$^a$Guangdong Provincial Key Laboratory of Turbulence Research and Applications, Department of Mechanics and Aerospace Engineering, Southern University of Science and Technology, Shenzhen 518055, PR China\\
$^b$Department of Mechanical and Aerospace Engineering, Hong Kong University of Science and Technology, Hong Kong, PR China}}}

\begin{abstract}
Reservoir computing is a powerful tool for forecasting turbulence because its simple architecture has the computational efficiency to handle large systems. Its implementation, however, often requires full state-vector measurements and knowledge of the system nonlinearities. We use nonlinear projector functions to expand the system measurements to a high dimensional space and then feed them to a reservoir to obtain forecasts. We demonstrate the application of such reservoir computing networks on spatiotemporally chaotic systems, which model several features of turbulence. We show that using radial basis functions as nonlinear projectors enables complex system nonlinearities to be captured robustly even with only partial observations and without knowing the governing equations. Finally, we show that when measurements are sparse or incomplete and noisy, such that even the governing equations become inaccurate, our networks can still produce reasonably accurate forecasts, thus paving the way towards model-free forecasting of practical turbulent systems.
\end{abstract}



\end{frontmatter}


\section{Introduction}
\label{Int}
The availability of large datasets and advances in computational hardware have led to revolutionary applications of machine learning (ML), such as in speech recognition~\citep{Hinton2012} and medical diagnosis~\citep{Alber2019}.
This has led to renewed interest in the application of ML to predict the behaviour of turbulent flow systems, such as for weather and climate forecasting~\citep{Shi2017, Wan2017, Wan2018, Reichstein2019, Kashinath2021, Watt-Meyer2021, Pathak2022}.
Turbulent systems are defined by processes spanning a wide range of spatial and temporal scales. These processes are coupled nonlinearly such that even small changes at the microscopic scale can lead to large changes at the macroscopic scale~\citep{Leith1972}, a phenomenon poetically termed the ``butterfly effect'' by Lorenz~\citep{Lorenz}.
Modelling such nonlinearly coupled dynamical systems with recurrent neural networks (RNNs) would require the network dimension to be exceedingly large, limiting their applicability to either coarse predictions~\citep{Shi2017, Wan2017} or necessitating dimensionality reduction of the input states~\citep{Vlachas2018, Li2020, Vlachas2022}.

Reservoir-computing-based RNNs (RC-RNNs) have been shown to be more computationally efficient at learning large chaotic systems as compared to backpropagation RNNs~\citep{Chattopadhyay2020, Vlachas2020}. Their implementation, however, often requires knowledge of the type of nonlinearities in the governing equations~\citep{Pathak2018, Chattopadhyay2020}.
This hinders the application of RC-RNNs to natural and engineering systems where full state-vector measurements are often not available and/or the governing equations are not known.
In this paper, we introduce a modification to RC-RNNs whereby nonlinear projectors are used to first expand the observables to a high dimensional space before they are fed to the reservoir.
We demonstrate the application of such networks for forecasting spatiotemporally chaotic systems, which are nonlinear dynamical systems with a large number of degrees of freedom that can model several important features of turbulence~\citep{Yakhot1981, Cross1993}.
In particular, we explore the use of radial basis functions (RBFs) as nonlinear projectors to enable the networks to capture complex system nonlinearities without knowing the governing equations.
We then explore scenarios in which only sparse or noisy and incomplete measurements are available, and compare the predictions from RC-RNNs with those from conventional time-marching.

\section{Network details}
\label{Net}

\begin{figure*}
 \centerline{\includegraphics[width=0.85\textwidth, trim =2cm 1.05cm 14.2cm 0.2cm, clip]{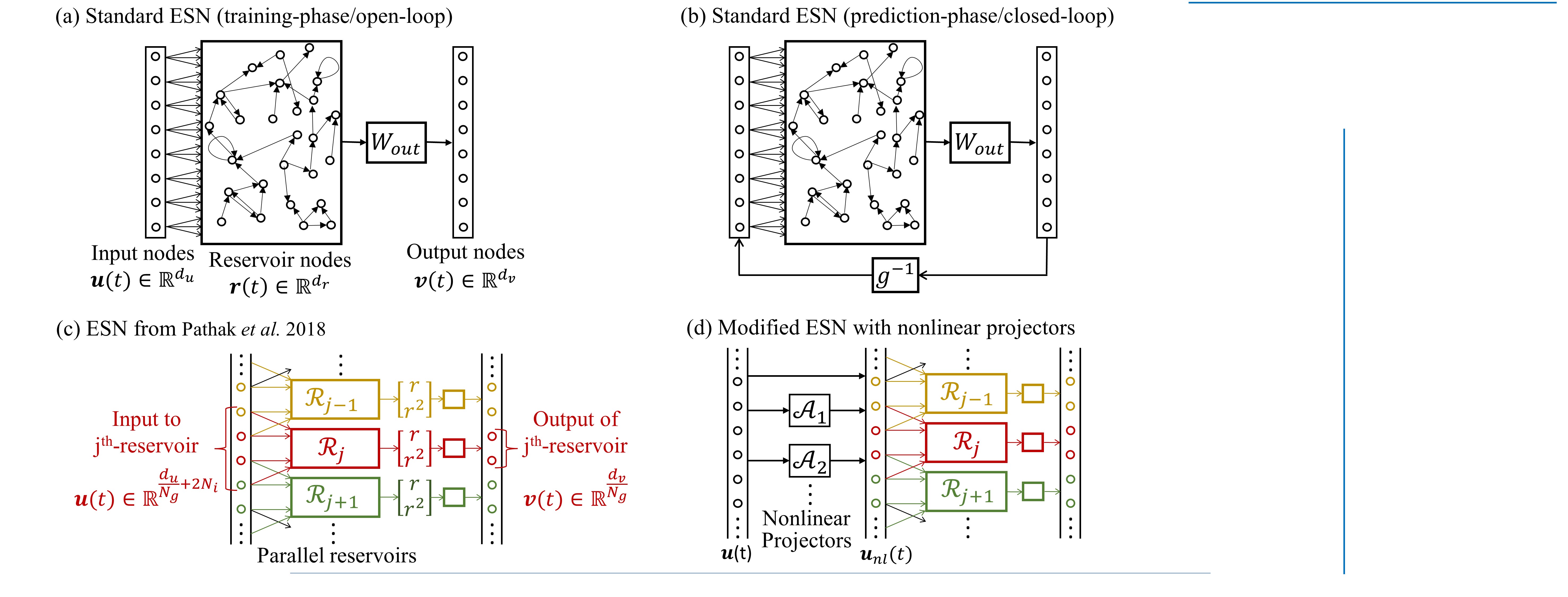}}
  \caption{(a)  Schematic of a standard ESN. The input nodes are connected to a high-dimensional reservoir, which is the only hidden layer in the network. A large enough reservoir can capture the system nonlinearities from which the output (future predictions here) is obtained via linear readout. (b) The network in the prediction phase is closed-loop, i.e. the output of the network becomes its input at the next step. (c) Schematic of the ESN from \citep{Pathak2018}; it has two improvements: (i) it is parallelized and (ii) the reservoir state is expanded to a higher-dimension (such as from $r$ to $[r,r^2]$). (d) Schematic of the proposed modified ESN; the observed vector $\mathbf{u}(t)$ is expanded to $\mathbf{u}_{nl}(t)$ via nonlinear projectors $\mathscr{A}_1$, $\mathscr{A}_2$, etc (i.e. $u_{nl}(t) = [u(t), \mathscr{A}_1(u(t)), \mathscr{A}_2(u(t)), ...]$. This means that the nonlinearity in the modified network is introduced before the reservoir instead of after it as in the network in (c).}
\label{RCs}
\end{figure*}

We choose the echo-state network (ESN) of Jaeger~\citep{Jaeger2004} for its ability to forecast chaotic systems. However, we emphasize that our work is applicable to other RNNs as well.
In ESNs the input state $\mathbf{u}(t)\in \mathbb{R}^{d_u}$, which consists of $d_u$ concurrent measurements in space, is connected to a high-dimensional dynamical system, say $\mathcal{R}$, known as the reservoir. The reservoir has the activation state $\mathbf{r}(t) \in \mathbb{R}^{d_r}$, which is effectively a nonlinear function of the input history $\mathbf{u}(t), \mathbf{u}(t-\Delta t), \dots$, where $\Delta t$ is a discrete time step (Eq.~\ref{res1}). The output $\mathbf{v}(t) \in \mathbb{R}^{d_v}$ is then obtained as a linear readout from $\mathbf{r}(t)$ using  Eq.~\ref{out1}.
\begin{subequations}
\begin{equation}\label{res1}
\mathbf{r}(t) = \mathcal{N}\left( A\mathbf{r}(t-\Delta t) + W_{in}\mathbf{u}(t)\right),
\end{equation}\\[-30pt]
\begin{equation}\label{out1}
\mathbf{v}(t) = W_{out}\mathbf{r}(t).
\end{equation}
\end{subequations}
Here, $W_{in} \in \mathbb{R}^{d_r\times d_u}$ connects the input state $\mathbf{u}(t)$ to the reservoir nodes, $A \in \mathbb{R}^{d_r\times d_r}$ is an adjacency matrix that interconnects the reservoir nodes, $\mathcal{N}$ is a known nonlinear function ($\tanh$ here) that acts on each element separately (i.e. $\mathcal{N}(\left[a_1, a_2, \dots\right]) = \left[\mathcal{N}(a_1),\mathcal{N}(a_2), \dots\right]$) and $W_{out} \in \mathbb{R}^{d_v\times d_r}$ connects the reservoir nodes to the output  (see the schematic in Fig.~\ref{RCs} (a)).

The elements of $W_{in}$ and $A$ are fixed by drawing them from random distributions, such as from a uniform distribution in $[-1, 1]$.
The elements of $A$ are scaled to keep its spectral radius ($\rho$) usually below one so that the network satisfies the fading memory property, i.e. the future state depends only on a finite sequence of past inputs.
Finally, the elements of $W_{out}$ are optimized via a one-step linear regression that minimizes the mean-square error between $\mathbf{v}(t)$ and the target signal $\widehat{\mathbf{v}}(t)$ during the training time $-T \leq t \leq 0$.
In forecasting problems, which are of interest here, the target signal $\widehat{\mathbf{v}}(t) = g(\mathbf{u}(t+1))$, where $g$ is a bijective function. Consequently, the network becomes closed-loop in the prediction phase, i.e. the model output becomes its input at the next step as shown in Fig.~\ref{RCs} (b).
See~\ref{A1} for further details and~\citep{Lukosevicius2012} for practical guidance on designing ESNs.


If the complex nonlinearities of turbulent systems are to be captured, $d_r$ may need to be several orders of magnitude larger than $d_u$. This can be computationally costly for turbulent systems for which $d_u$ is already large.
Recognizing this, Pathak \textit{et al.} (2018) \citep{Pathak2018} proposed two improvements (see Fig.~\ref{RCs} (c)). First, they parallelized the ESNs based on the spatial locality of turbulent interactions, which facilitates parallel computing. The parallelized network has two additional hyperparameters: the number of parallel reservoirs $N_g$ and the interaction length $N_i$. Second, they explicitly expanded the reservoir activation state to a higher-dimensional space such that the output is obtained as $\mathbf{v}(t) = W_{out} \left[\mathbf{r}, \mathbf{r}^2\right]$. The main drawback of this approach is that in order to capture the system nonlinearities, the nonlinearity added explicitly should closely match that present in the system equations themselves~\citep{Chattopadhyay2020}.

We will show that the network from Pathak \textit{et al.} (2018) \citep{Pathak2018} can be improved if the expansion to the high-dimensional space happens before the reservoir. In other words, we expand the observables (i.e. $u \in \mathbb{R}^{d_u}$) to a high-dimensional space using nonlinear projector functions  $\mathscr{A}_{l}$ ($l = 1, 2, ..., L$). The expanded vector $u_{nl} \in \mathbb{R}^{(L+1)d_u}$ is then the input to the reservoir from which the output is trained via linear readout (see Fig.~\ref{RCs} (d)).

\section{Predictions from existing and modified networks}
\label{Res1}

\begin{figure*}
 \centerline{\includegraphics[width=1.0\textwidth, trim =1.1cm 4.65cm 6.9cm 0.25cm, clip]{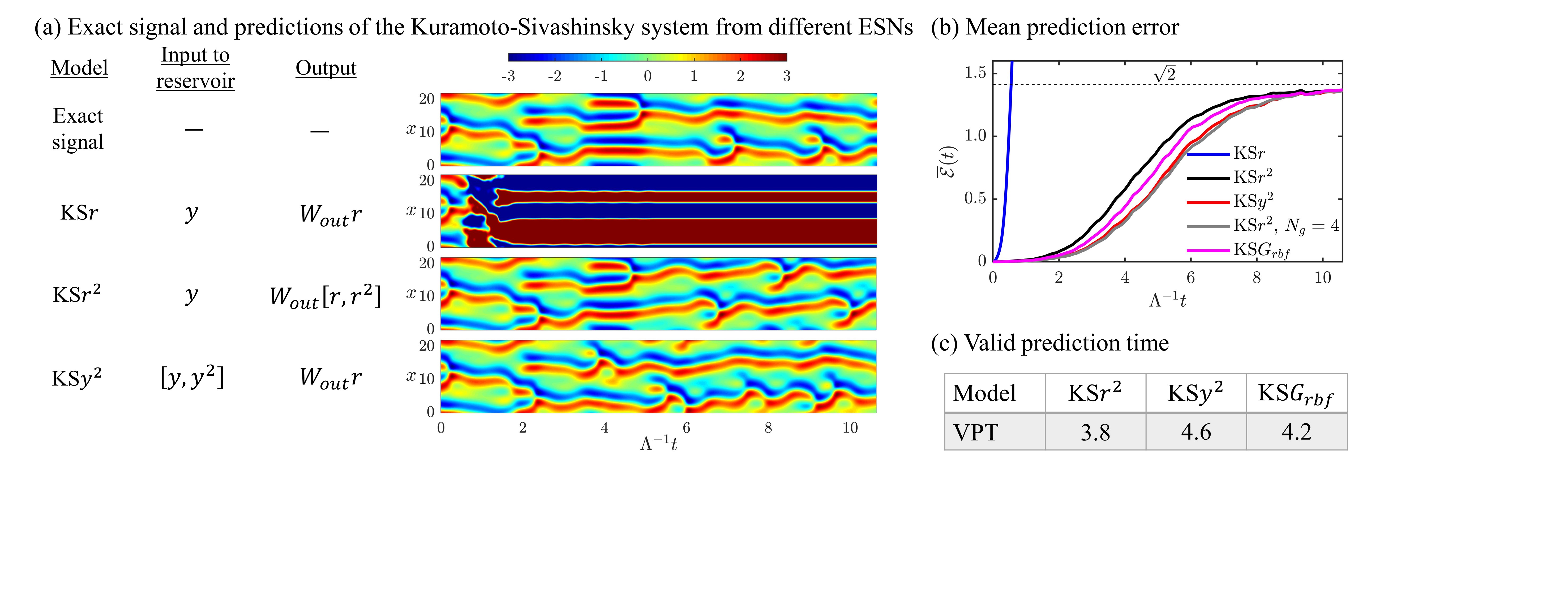}}
  \caption{(a) The exact signal for the KS system and its predictions from different RC-RNNs: KS$r$ (an instance of a standard network shown in Fig.~\ref{RCs} (a)), KS$r^2$ (an instance of the network from \citep{Pathak2018} shown in Fig.~\ref{RCs} (c)) and KS$y^2$ (an instance of the proposed network shown in Fig.~\ref{RCs} (d)). All networks here are of size $d_r = 1000$ and $N_g=1$ (i.e. the reservoir size is only an order of magnitude larger than $d_u$). The KS$r$ network, which has no explicitly added nonlinearity, cannot capture the system nonlinearities (unless maybe for much larger $d_r$). The networks with added square nonlinearities (KS$r^2$ and KS$y^2$) are able to capture the system nonlinearities well. The mean prediction errors (b) and the valid prediction time (c) from the different networks are also shown. The modified network KS$y^2$ has relatively slower growth of error as compared to the network of \citep{Pathak2018} (KS$r^2$). The improvement is significant considering that the KS$r^2$ network would require four times more computational power to achieve similar performance to the KS$y^2$ network. The KS$G_{rbf}$ network is also an instance of the proposed network but with input to reservoir as  $[y,\exp (-0.03y^2)]$, this network is discussed with Fig.~\ref{Efit} (d-f).}
\label{KSpreds}
\end{figure*}

We consider the Kuramoto--Sivashinsky (KS) equation as our first example:
\begin{equation}\label{ks}
y_t  = -yy_x - y_{xx} - \nu y_{xxxx}, \;\;\;\;\;\; x \in [0,L].
\end{equation}
The scalar field $y(x,t)$ is periodic in space. The subscripts $x$ and $t$ denote the partial derivatives in space and time, respectively. The parameter $\nu$ acts like the viscosity in the Navier--Stokes equations~\citep{Bratanov2013} and $L$ is the system size with which the attractor dimension increases linearly~\citep{Manneville1985}. 
We numerically solve Eq.~\ref{ks} for $(\nu,L) = (1,22)$ with $N=128$ gridpoints and timestep $dt = 0.25$~\citep{Kassam2005}. In this study $\Delta t$ of networks is set as $dt$.

Fig.~\ref{KSpreds} (a) shows the exact signal from the simulations and its prediction from different ESNs. All the networks feature a single reservoir of size $d_r = 1000$, which is only an order of magnitude larger than $d_u = 128$. The networks are trained for $-10000 \leq t \leq 0$, training with much smaller time history is possible but the predictive accuracy will then be lower as shown in~\citep{Pathak2018}. We note that the KS$r$ network, which is an instance of the standard network (Fig.~\ref{RCs} (a)), is unable to learn the system dynamics, with the corresponding forecast soon becoming uninterpretable. Nevertheless, the results drastically improve when the networks with added nonlinearities are used. Here KS$r^2$ and KS$y^2$ are instances of the network from~\citep{Pathak2018} (Fig.~\ref{RCs} (c)) and the proposed network (Fig.~\ref{RCs} (d)), respectively. Both networks predict the exact signal in the short-term after which the patterns in the predictions appear to be spatiotemporally shifted. This shift is an inherent property of chaotic systems: their trajectories diverge exponentially in time even when they start from infinitesimally close initial conditions. The rate of separation is given by $\exp(\Lambda_{max}t)$, where $\Lambda_{max}$ is the maximal Lyapunov exponent and is positive for chaotic systems~\citep{Boffetta2002}.

 %

We quantify the prediction error as $\mathcal{E}(t) = \sqrt{\langle(\mathbf{v}(t) - \widehat{\mathbf{v}}(t))^2\rangle}/\sigma_y$,
where $\langle.\rangle$ denotes spatial averaging and $\sigma_y$ is the standard deviation of $y(x,t)$.
To account for variability when comparing the performance of different ESNs, we take the mean of the error ($\overline{\mathcal{E}(t)}$) for a series of predictions (see~\ref{A2} for details). Fig.~ref{KSpreds} (b) shows $\overline{\mathcal{E}(t)}$ for predictions from various ESNs. The mean error increases exponentially from any infinitesimal differences in the predictions at $t = 0$. 
A slower increase in $\overline{\mathcal{E}(t)}$ would reduce the effect of both measurement and model uncertainties on future predictions.
However, we note that the mean-square-error contains only limited information~\citep{Majda2018}.
For the predictions to be physically interpretable, their probability distribution should be close to that of the exact signal~\citep{DelSole2004}. (This is the reason for the popularity of generative-adversarial networks~\citep{Goodfellow2014}, which can give better interpretable results often at the cost of increased mean-square-error~\citep{Kim2021}.)
If the predictions are physically interpretable in the long-term, $\overline{\mathcal{E}(t)}$ would saturate to a value close to $\sqrt{2}$.

As expected from the predictions in Fig.~\ref{KSpreds} (a), $\overline{\mathcal{E}(t)}$ from the standard network KS$r$ not only increases rapidly but it also does not saturate to $\sqrt{2}$, i.e. the predictions are not interpretable. For all other networks used here, $\overline{\mathcal{E}(t)}$ saturates to $\sqrt{2}$ thus indicating statistically consistent predictions. We note that the modified network KS$y^2$ has smaller growth-rate of error as compared to that for the network from \citep{Pathak2018} (KS$r^2$). In fact, the KS$r^2$ network requires almost four times more computational power (i.e. $N_g =4$) to achieve a similar level of accuracy as the KS$y^2$ network. Although both networks have explicitly added square nonlinearities, the KS$y^2$ network outperforms the KS$r^2$ network. This is because in the KS$y^2$ network the reservoir receives more information about the system. In other words, the reservoir in the KS$y^2$ network is preconditioned with square nonlinearity and can thus better approximate the system nonlinearity.

Another metric recently introduced for evaluating the predictive performance of models is the valid prediction time (VPT) \citep{Vlachas2020}. It refers to the time at which $\overline{\mathcal{E}(t)}$ crosses a threshold; the threshold is set as 0.5 following \citep{Vlachas2020}. The advantage of using VPT is that it is a scalar measure that quantifies the predictive accuracy of different methods. Networks with slower growth-rate of the mean-square error also have larger VPT as shown in Fig.~\ref{KSpreds} (c) for the KS system; this trend is independent of the threshold value.

\section{Partially observable systems: selection of nonlinear projectors}\label{R2}

\begin{figure*}
 \centerline{\includegraphics[width=1\textwidth, trim =1.1cm 1.95cm 6.3cm 0.25cm, clip]{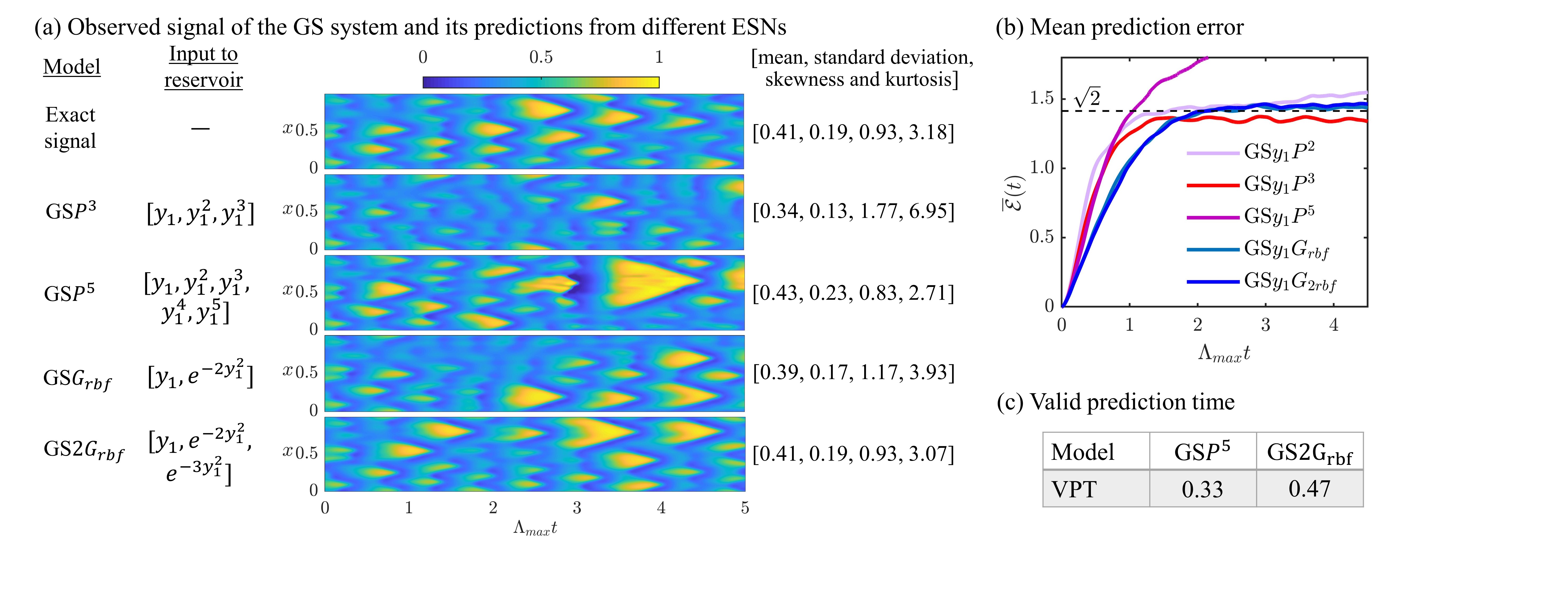}}
  \caption{(a) The observed signal, $y_1$, of the GS system and its predictions from the GS$P^3$, GS$P^5$, GS$G_{rbf}$ and GS$2G_{rbf}$ networks of size $d_r = 1500$ and $N_g=4$ (i.e. the total reservoir size is 6000, which is only an order of magnitude larger than $d_u$). Although all four networks can predict the short-term dynamics, the networks with monomial projectors (GS$P^3$ and GS$P^5$) fail to give statically consistent results while the networks with Gaussian RBF projectors (GS$G_{rbf}$ and GS$2G_{rbf}$) give statistically consistent results. Also shown are the mean prediction errors (b) and the valid prediction time (c) from the different ESNs. In addition to statistically consistent results, the networks with Gaussian RBF projectors also show slower growth-rates of the error and hence higher VPTs.}
\label{GSpreds}
\end{figure*}

For the KS system, we showed that the networks with explicitly added square nonlinearities (KS$r^2$ and KS$y^2$) work well. As noted in \citep{Chattopadhyay2020}, this is because the added nonlinearity in the networks matches the nonlinearity in the system equations (i.e. square nonlinearity in the KS system here). The form of system nonlinearity, however, is not always known; for example the governing equations are not known or full state-vector measurements are not available.
To illustrate, we consider the Gray--Scott (GS) system, which models  the reaction-diffusion of two species, as our second example,
\begin{subequations}\label{gs}
\begin{equation}\label{gs1}
y_{1t} = \left(F - y_1y_2^2\right) - Fy_1 + D_1 y_{1xx},
\end{equation}\\[-30pt]
\begin{equation}\label{gs2}
y_{2t} = y_1y_2^2 - (F + k)y_2 + D_2 y_{2xx},
\end{equation}
\end{subequations}
where $y_1(x,t)$ and $y_2(x,t)$ are the concentration of two chemical species. 
We choose $D_1 = 2.0e^{-5}$, $D_2 = 1.0e^{-5}$, $F = 0.015$, $k = 0.0415$ and $x \in [0,1]$ with periodic boundary conditions for which the system exhibits annihilation type spatiotemporal chaos~\citep{Nishiura2001}. We numerically solve the GS system with 200 chebyshev points and $dt = 0.5$  and save the results on an equispaced grid containing 256 nodes~\citep{Driscoll2014}. The time evolution of the concentration of species one ($y_1$) is shown in Fig.~\ref{GSpreds} (a). 

We now assume that only $y_1$ measurements are available ($\mathbf{u}(t) = y_1(x,t)$), i.e. we do not know the full state-vector.
One may argue that we can still form the full state-vector in terms of $(y_1, y_{1t})$ or calculate $y_2$ using $(y_1, y_{1t}, y_{1xx})$. This will, however, require knowledge of Eq.~\ref{gs} and noise free measurements of  $y_1$ with high spatiotemporal resolution. Neither are usually available in practical situations, nor are they needed by our method (see the results in Figs.~\ref{KSsp} and~\ref{GSns}).

With only $y_1$ measurements available, the form of system nonlinearity is not known. The challenge is therefore to select nonlinear projectors $\mathscr{A}_l$ that can enable the network to capture the nonlinear dynamics of the GS system from $y_1$ measurements alone.
First we try monomial projectors, such as $y_1^2$, $y_1^3$, etc., as used in new-generation-ESNs~\citep{Gauthier2021} and often used to approximate nonlinearity in turbulent systems~\citep{Brunton2016}.
(GS$P^n$ refers to an ESN with nonlinear projectors $\mathscr{A}_l = y_1^{l+1}, \; l = 1,2,..., n-1$.)
We note from Fig.~\ref{GSpreds} (b) that the growth of $\overline{\mathcal{E}}$ slows down slightly with increasing number of monomials but $\overline{\mathcal{E}}$ does not saturate to $\sqrt{2}$ for any of the GS$P^n$ networks.
Fig.~\ref{GSpreds} (a) shows poor quality of the predictions from the GS$P^n$ networks. The statistics here is calculated based on the closed-loop predictions from $0 \leq t \leq 25000$. We find that adding more monomials or reservoir units does not significantly improve the quality of predictions from the GS$P^n$ networks (results not shown).

Next, we try Gaussian RBFs as nonlinear projectors. This choice is inspired by the robust approximation properties of RBFs~\citep{Buhmann2000} on which RBF networks are based~\citep{Girosi1990, Zemouri2003}. For each Gaussian RBF projector $\mathscr{A} = \exp\left(-\gamma\left(y_1(x,t)-y_c(x)\right)^2\right)$, there are two parameters: (i) a shape factor $\gamma$ and (ii) a space-dependent centre $y_c$. As in RBF networks, we choose the centre in an unsupervised manner and optimize for $\gamma$ via a trial-and-correction procedure. In the GS system, $y_1(x,t)$  varies from 0 to 1, and the degree of nonlinearity increases from 1 to 0. We therefore choose $y_c(x) = 0$ as the centre ($y_c(x) = 1$ is an equally good choice). Fig.~\ref{GSpreds} shows that the GS$G_{rbf}$ network, which uses a Gaussian RBF projector $\exp\left(-2y_1^2\right)$, performs noticeably better, both in terms of short-term accuracy as well as the statistics of the predictions, as compared to any of the GS$P^n$ networks. The results are further improved, particularly in terms of the statistics, when two Gaussian RBF projectors ($\exp\left(-2y_1^2\right)$ and $\exp\left(-3y_1^2\right)$) are used in the GS$2G_{rbf}$ network.

Fig.~\ref{GSpreds} (c) shows the VPT from the GS$P^5$ and GS$2G_{rbf}$ networks, which highlights the significant improvement in the predictive accuracy when Gaussian RBFs are used instead of monomials as nonlinear projectors. The small VPT values from all the networks indicate the difficulty in obtaining accurate predictions from partial observations. This difficulty can be further appreciated from Fig.~\ref{GSns} in which the governing equations are used for obtaining the predictions.

\section{Relevance of nonlinear approximation properties of nonlinear projectors}\label{R3}

\begin{figure}
 \centerline{\includegraphics[width=1\textwidth, trim =1.6cm 8.85cm 8.25cm 0.2cm, clip]{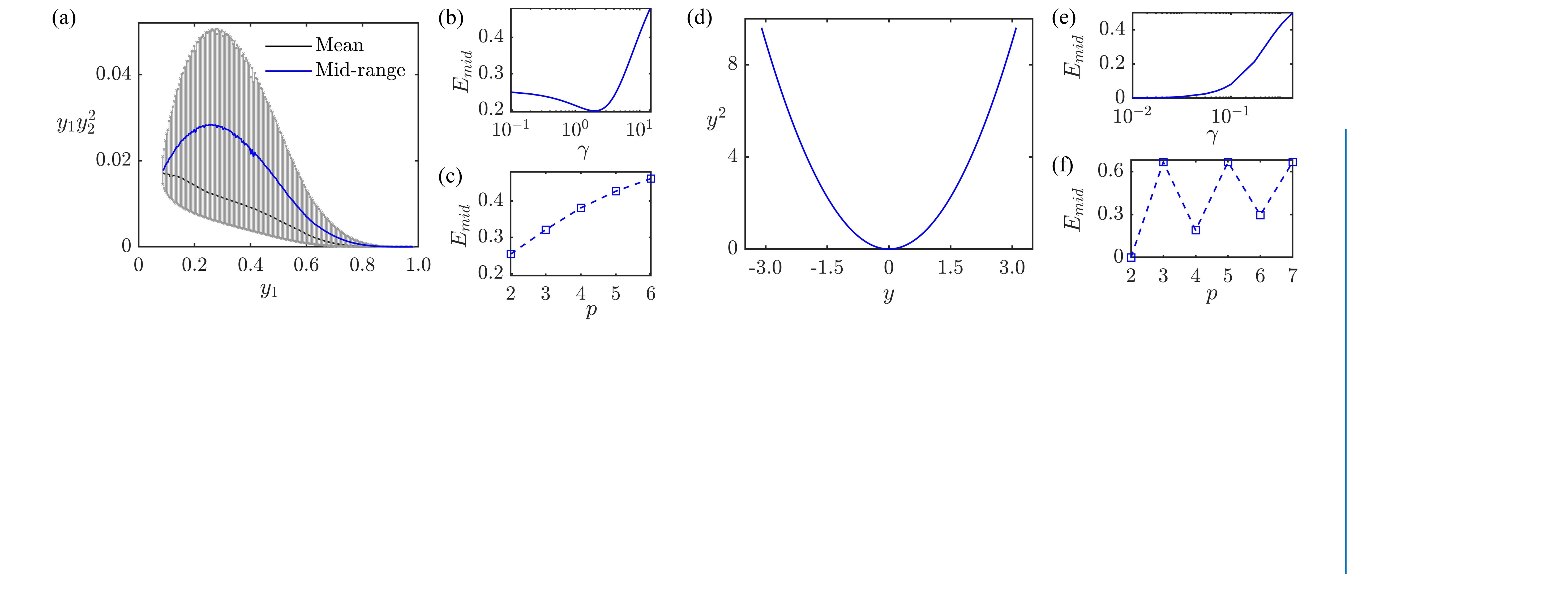}}
\caption{Nonlinear terms (a, d) and errors in approximating their mid-range values via Gaussian RBFs (b, e) and monomials (c, f) for the GS system (a-c) and the KS system (d-f). For the GS system, the Gaussian RBFs have relatively lower error, thus explaining the superior performance of the networks with the Gaussian RBF projectors. For the KS system, even when the nonlinearity is a monomial, the Gaussian RBF can approximate the nonlinear term well.
}
\label{Efit}
\end{figure}

We explore the connection between the nonlinear approximation properties of the projector functions and the performance of the corresponding ESNs.
Fig.~\ref{Efit} (a) shows the variation of the nonlinear term $y_1y_2^2(y_1)$, its mean and mid-range values ($m$) with the observed variable $y_1$ in the GS system. Here $m$ is the mean of the maximum and minimum values of $y_1y_2^2$ at a given $y_1$.
The error in approximating $m$ by a nonlinear projector function ($\mathscr{A}$) is given as
\begin{equation}
E_{mid} = \min_{a_1, a_2}\sqrt{\langle\left(a_1 +a_2\mathscr{A} - m\right)^2\rangle}/\sqrt{\langle m^2\rangle},
\end{equation}
where $\mathscr{A} = \exp(-\gamma y_1^2)$ for Gaussian RBFs (the corresponding $E_{mid}$ is shown in subfigure (b)) and $\mathscr{A} = y_1^p$ for monomials (the corresponding $E_{mid}$ is shown in subfigure (c)).
There are three points to note here. First, the value of $\gamma$ at which GS$G_{rbf}$ performs best coincides with that at which $E_{mid}$ is minimum (i.e. $\gamma \approx 2$), demonstrating that a good nonlinear projector does indeed approximate the system nonlinearity. Second, minimization of the error with respect to the mid-range value (rather than the mean value) indicates that a good nonlinear projector minimizes the chances of maximum error (i.e. $L_{\infty}$-error)~\citep{Giloni2002}. Third, $E_{mid}$ is lower for the Gaussian RBFs than for the monomials, thus explaining the superior performance of the GS$G_{rbf}$ and GS$2G_{rbf}$ networks over the GS$P^n$ networks in Fig.~\ref{GSpreds}. 
Such statistical interpretation indicates that RBFs can be useful in other RNNs as well.

Furthermore, the robustness of RBFs is evidenced by their ability to satisfactorily approximate most types of nonlinearities. For example, $\mathscr{A} = y^2$ is the best projector for the KS system for which the nonlinear term itself is $y^2$. However, Gaussian RBFs can still approximate $y^2$ nonlinearity well (see Fig~\ref{Efit} (e)) and the corresponding ESN (KS$G_{rbf}$) predicts the KS system relatively well (see Fig.~ref{KSpreds} (b-c)). 
Here we have only shown the results for Gaussian RBFs; the results are similarly good when multiquadric RBFs are used (see~\ref{A3}).

\section{Comparison between model-free and governing-equations-based predictions}\label{R4}

\subsection{Application to Kuramoto--Sivashinsky system with sparse measurements}\label{Sp}

\begin{figure*}
 \centerline{\includegraphics[width=1\textwidth, trim =0.1cm 4.15cm 0.3cm 0.5cm, clip]{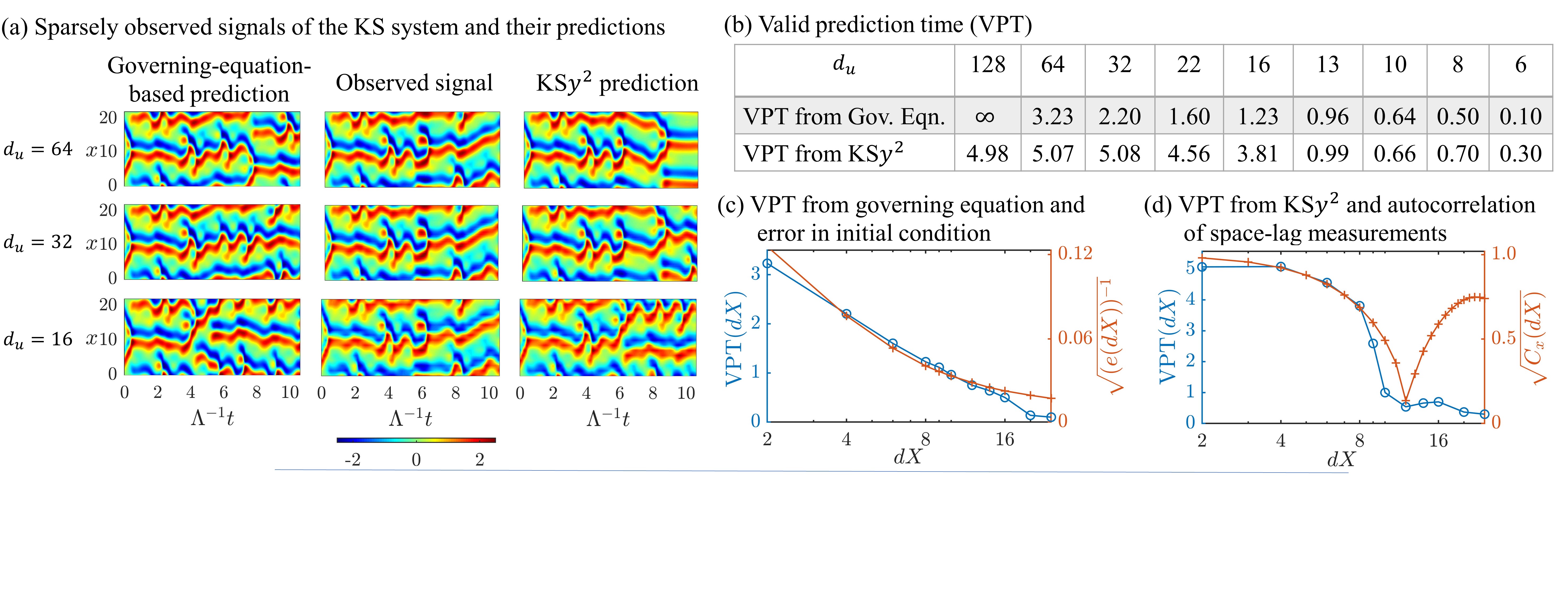}}
  \caption{(a) Sparsely observed signals of the KS system and their predictions from the governing equations and the KS$y^2$ network ($d_r = 2000$ and $N_g=1$). (b) The VPT of predictions at different levels of sparsity. The observed signal becomes increasingly sparse as $d_u$ decreases (or $dX \approx 128/d_u$ increases). (c) The VPT for the governing-equations-based predictions decreases rapidly with increasing $dX$, varying with the inverse of the square-root of the root-mean-square-error in the initial condition ($e(dX)$). (d) The VPT for the KS$y^2$ predictions are relatively unaffected as long as the autocorrelation between observations at neighbouring spatial locations ($C_x(dX)$) remains close to one.}
\label{KSsp}
\end{figure*}

We now consider a situation where only sparse measurements of the KS system are available. We compare the predictive performance of the governing-equations-based time-marching with that of the KS$y^2$ network. We take the ground truth as the original simulation with $N =128$ gridpoints. The observed signal ($y_{obs}(x,t)$), shown in the middle column of Fig.~\ref{KSsp} (a), consists of measurements at $d_u$ equispaced points at distance $dX L/N$. This implies that the observations become increasingly sparse with decreasing $d_u$ (or increasing $dX$).

The governing-equations-based predictions are obtained by time-marching from the initial condition at $t=0$. The initial condition, say $y_{int}(x)$, is obtained by linearly interpolating $y_{obs}(x,t=0)$ from the observed space (consisting of $d_u$ points) to the numerical grid (consisting of $N=128$ points). The predictions and the VPT for different levels of sparsity in the observations are shown in Fig.~\ref{KSsp} (a) and (b), respectively. The accuracy of the predictions decreases with increasing sparsity as indicated by the reduction in VPT. Fig.~\ref{KSsp}~(c) shows that the reduction in VPT is proportional to $\sqrt{e(dX)^{-1}}$, where $e(dX) = \sqrt{\langle \left(y(x,t=0) - y_{int}(x,t=0)\right)^2\rangle}$ is the root-mean-square error in the initial condition. We note that the error in the initial conditions could be reduced by using other interpolation schemes or data assimilation~\citep{Whitaker2009, Brajard2020}. However, the trend would remain the same because an interpolation scheme cannot entirely compensate for the missing information at smaller scales. We also note that the system parameters and boundary conditions are known perfectly here, which often contribute to error in the governing-equations-based predictions.

The predictions from the KS$y^2$ network are obtained by training the network with the sparsely observed data, i.e. the network never sees the original data. The predictions and VPT from the KS$y^2$ network for different levels of sparsity in the observations are shown in Fig.~\ref{KSsp} (a) and (b), respectively.
We find that the predictions from the RC-RNN are insensitive to the discretization or interpolation error; a similar level of accuracy is maintained even when $d_u$ is reduced from $N$ to $N/4$. Fig.~\ref{KSsp} (d) shows that VPT from KS$y^2$ is proportional to $\sqrt{C_x(dX)}$, where $C_x(dX)$ is the mean autocorrelation between $y(x)$ and $y(x+dXL/N)$ signals. The results show that information about the well correlated elements of the state-vector can be learned from the time history of partial measurements. This suggests that RC-RNNs can unravel the space-time dynamics in agreement with Takens' embedding theorem~\citep{Takens1981}, which has recently been suggested to be the case also for data-driven discovery of the governing equations from partial measurements~\citep{Bakarji2022}.

The results show that RC-RNNs can provide excellent predictions up to moderate levels of sparsity even when the governing equations may suffer from high discretization errors. The accuracy of both methods can be improved, such as by using better interpolation schemes for the governing equations or by increasing the reservoir dimension for the RC-RNN. The key point, however, is that RC-RNNs are less affected by the sparsity as compared to governing-equations-based time-marching.
When the sparsity is high such that the spatially adjacent measurements are beyond the first zero in the autocorrelation, we find that the RC-RNNs perform poorly. We expect the same limitations for other data-driven methods as well

\subsection{Application to Gray--Scott system with partial and noisy measurements}\label{Noisy}

\begin{figure}
 \centerline{\includegraphics[width=1\textwidth, trim =0cm 5.45cm 6.4cm 0cm, clip]{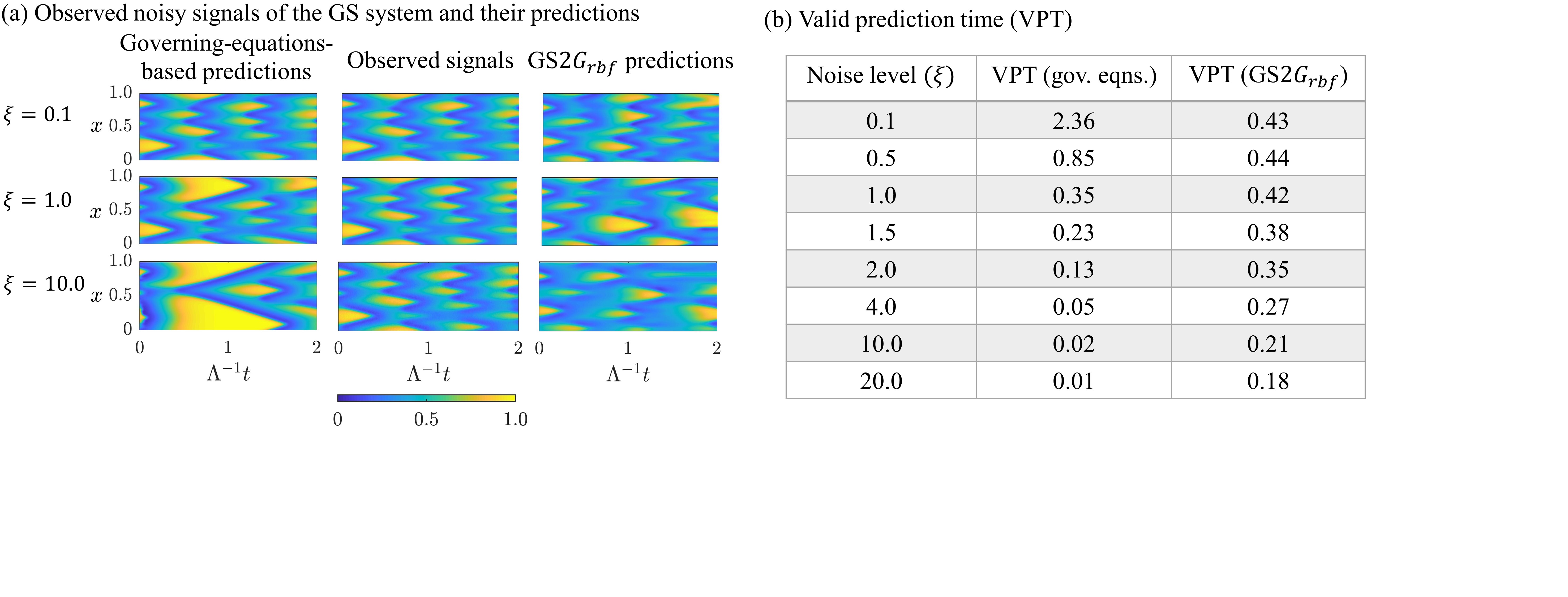}}
\caption{(a) Partially observed noisy signals for the GS system and their predictions from the governing equations and the GS$2G_{rbf}$ network (of size $d_r = 1500$ and $N_g=4$). (b) VPT for the predictions from the governing equations and the GS$2G_{rbf}$ network. A reduction in predictive accuracy occurs with increasing noise level. The GS$2G_{rbf}$ network, however, is relatively less affected by the noise. The small values of VPT from both methods demonstrate the challenge of forecasting chaotic systems with partial observability.}
\label{GSns}
\end{figure}

We now consider another practical scenario in which only partial and noisy observables are available. We take the ground truth as the original simulation of the GS system. The concentration of species one with measurement noise, say $y_{1obs}$, is the observable (shown in the middle column of Fig.~\ref{GSns}~(a)). The measurement noise is additive, white in space and time, with uniform distribution between $[-\xi \sigma_{y_1}, \xi \sigma_{y_1}]$, where $\sigma_{y_1}$ is the standard deviation of $y_1$ and $\xi$ is the noise level. Similar to the sparse measurement cases, the governing-equations-based predictions are obtained by time-marching from the initial conditions that are calculated from the noisy measurements of $y_1$ at $t=0$, while RC-RNN-based predictions are obtained by training the GS$2G_{rbf}$ network with the noisy measurements.

The governing equations, their parameters and the boundary conditions are all known perfectly, but the initial condition for $y_2$ is now estimated from the noisy measurements of $y_1$. This introduces high levels of error in the initial condition and, consequently, the accuracy of the governing-equations-based predictions drops drastically with increasing noise-level as shown in Fig.~\ref{GSns}.
The RC-RNNs are fundamentally different in that they predict the future by learning patterns in the time history. By design, therefore, they are robust to small levels of measurement noise. The results in Fig.~\ref{GSns} show that although the accuracy of the RC-RNN-based predictions also decrease with increasing noise level, they do not suffer as much as the governing-equations-based predictions.

\section{Conclusions}\label{Con}

The main conclusions of this paper are threefold.
The first two concern the explicitly added nonlinearity in RC-RNNs. Fig.~\ref{KSpreds} and Refs.~\citep{Pathak2018, Chattopadhyay2020} show it to be essential to explicitly add nonlinearities in RC-RNNs to enable them to capture complex system nonlinearities. Firstly, we find it more computationally efficient to add the nonlinearity before the reservoir, such as in the KS$y^2$ network, than after the reservoir, such as in the KS$r^2$ network. Adding the nonlinearity before the reservoir preconditions the network towards the system dynamics and hence approximate the system nonlinearities better. Secondly, we find that the use of RBFs as nonlinear projectors (to explicitly add nonlinearities in RC-RNNs) enables the networks to capture the system nonlinearities even with partial measurements and/or without knowing the governing equations.
We show that the robust approximation properties of RBFs, which makes them suitable for nonlinear approximation in numerical schemes and RBF networks~\citep{Girosi1990, Buhmann2000, Zemouri2003}, also makes them a good choice as nonlinear projectors in the RC-RNNs proposed here. The best nonlinear projectors approximate the mid-range value of the system nonlinearities, thus minimize the $L_{\infty}$-error. This statistical interpretation enhances the viability of using RBFs in other RNNs for turbulence prediction.

The third conclusion is on viability of RC-RNNs as an alternative to conventional methods in practical situations. Observations in natural and engineering systems can often have missing data, coarse resolution and noise; thus several ML methods have been developed to extract information from limited and noisy measurements~\citep{Reichstein2019, Raissi2020}. Prediction of turbulent systems, however, is extremely sensitive to any estimation errors. This is evident from the results in Figs.~\ref{KSsp} and~\ref{GSns}, which show that the governing equations, even with perfectly known boundary conditions and parameters, quickly lose predictive accuracy as the measurements become sparse, incomplete and noisy. We find that the predictive performance of RC-RNNs is relatively less affected by sparsity and noise in the measurements, thus can potentially outperform the governing equations in such cases.

In summary, we have demonstrated the capabilities of our proposed RC-RNN framework to forecast (i) partially observable spatiotemporally chaotic systems with no knowledge of the form of system nonlinearity and (ii) sparsely observable systems or partially observable systems with measurement noise for which even the governing equations become inaccurate.
No recourse to the governing equations is required, thus making our RC-RNN framework truly model-free. We emphasize that the partial and sparse observability in our work is fundamentally different from the model-reduction or partial observability discussed in recent studies on turbulence forecasting~\citep{Vlachas2018, Vlachas2022}. In those studies, the extraction of dominant system modes may require noise-free measurements of the full state-vector.
Finally, we also highlight the difficulty in predicting noisy and partially observable systems, indicating that much larger reservoirs than those used in this study may be required for useful predictions in real-world applications, such as weather forecasting. Understanding the relationship between the system dynamics and hyperparameters of RC-RNNs and the uncertainty quantification of the predictions would be vital to unlocking the full potential of applying RC-RNNs to practical turbulent systems.

\section{Acknowledgments}
We wish to thank Prof. Luca Biferale for his feedback during the preparation of this manuscript. This work was funded by the National Natural Science Foundation of China (Grant Nos. 12002147, 12050410247 and 11988102), the Shenzhen Science and Technology Program (Grant No. \\
KQTD20180411143441009 ) and the Department of Science and Technology of Guangdong Province (2020B1212030001). We acknowledge support from the Centers for Mechanical Engineering Research and Education at MIT and SUSTech, as well as from the Center for Computational Science and Engineering at SUSTech.. LLKB gratefully acknowledges the funding from the Research Grants Council of Hong Kong (Projects 16210419, 16200220, 16215521).

 \bibliographystyle{elsarticle-num} 
 \bibliography{Nref}

\begin{thebibliography}{10}
\expandafter\ifx\csname url\endcsname\relax
  \def\url#1{\texttt{#1}}\fi
\expandafter\ifx\csname urlprefix\endcsname\relax\def\urlprefix{URL }\fi
\expandafter\ifx\csname href\endcsname\relax
  \def\href#1#2{#2} \def\path#1{#1}\fi

\bibitem{Hinton2012}
G.~Hinton, L.~Deng, D.~Yu, G.~Dahl, A.~R. Mohamed, N.~Jaitly, A.~Senior,
  V.~Vanhoucke, P.~Nguyen, T.~Sainath, B.~Kingsbury, Deep neural networks for
  acoustic modeling in speech recognition: The shared views of four research
  groups, IEEE Signal Processing Magazine 29 (2012) 82--97.
\newblock \href {https://doi.org/10.1109/MSP.2012.2205597}
  {\path{doi:10.1109/MSP.2012.2205597}}.

\bibitem{Alber2019}
M.~Alber, A.~B. Tepole, W.~R. Cannon, S.~De, S.~Dura-Bernal, K.~Garikipati,
  G.~Karniadakis, W.~W. Lytton, P.~Perdikaris, L.~Petzold, E.~Kuhl, Integrating
  machine learning and multiscale modeling—perspectives, challenges, and
  opportunities in the biological, biomedical, and behavioral sciences, Digital
  Medicine 2 (12 2019).
\newblock \href {https://doi.org/10.1038/s41746-019-0193-y}
  {\path{doi:10.1038/s41746-019-0193-y}}.

\bibitem{Shi2017}
X.~Shi, Z.~Gao, L.~Lausen, H.~Wang, D.-Y. Yeung, W.-K. Wong, W.-C. Woo,
  \href{http://arxiv.org/abs/1706.03458}{Deep learning for precipitation
  nowcasting: A benchmark and a new model}, 2017, pp. 5622--5632.
\newline\urlprefix\url{http://arxiv.org/abs/1706.03458}

\bibitem{Wan2017}
Z.~Y. Wan, T.~P. Sapsis, \href{http://arxiv.org/abs/1611.01583
  http://dx.doi.org/10.1016/j.physd.2016.12.005}{Reduced-space gaussian process
  regression for data-driven probabilistic forecast of chaotic dynamical
  systems}, Physica D 345 (2017) 40--55.
\newblock \href {https://doi.org/10.1016/j.physd.2016.12.005}
  {\path{doi:10.1016/j.physd.2016.12.005}}.
\newline\urlprefix\url{http://arxiv.org/abs/1611.01583
  http://dx.doi.org/10.1016/j.physd.2016.12.005}

\bibitem{Wan2018}
Z.~Y. Wan, P.~Vlachas, P.~Koumoutsakos, T.~Sapsis, Data-assisted reduced-order
  modeling of extreme events in complex dynamical systems, PLoS ONE 13 (5
  2018).
\newblock \href {https://doi.org/10.1371/journal.pone.0197704}
  {\path{doi:10.1371/journal.pone.0197704}}.

\bibitem{Reichstein2019}
M.~Reichstein, G.~Camps-Valls, B.~Stevens, M.~Jung, J.~Denzler, N.~Carvalhais,
  Prabhat, Deep learning and process understanding for data-driven earth system
  science, Nature 566 (2019) 195--204.
\newblock \href {https://doi.org/10.1038/s41586-019-0912-1}
  {\path{doi:10.1038/s41586-019-0912-1}}.

\bibitem{Kashinath2021}
K.~Kashinath, M.~Mustafa, A.~Albert, J.~L. Wu, C.~Jiang, S.~Esmaeilzadeh,
  K.~Azizzadenesheli, R.~Wang, A.~Chattopadhyay, A.~Singh, A.~Manepalli,
  D.~Chirila, R.~Yu, R.~Walters, B.~White, H.~Xiao, H.~A. Tchelepi, P.~Marcus,
  A.~Anandkumar, P.~Hassanzadeh, Prabhat, Physics-informed machine learning:
  Case studies for weather and climate modelling, Phil. Trans. R. Soc. A 379
  (2021) 20200093.
\newblock \href {https://doi.org/10.1098/rsta.2020.0093}
  {\path{doi:10.1098/rsta.2020.0093}}.

\bibitem{Watt-Meyer2021}
O.~Watt-Meyer, N.~D. Brenowitz, S.~K. Clark, B.~Henn, A.~Kwa, J.~McGibbon,
  W.~A. Perkins, C.~S. Bretherton, Correcting weather and climate models by
  machine learning nudged historical simulations, Geophysical Research Letters
  48 (8 2021).
\newblock \href {https://doi.org/10.1029/2021GL092555}
  {\path{doi:10.1029/2021GL092555}}.

\bibitem{Pathak2022}
J.~Pathak, S.~Subramanian, P.~Harrington, S.~Raja, A.~Chattopadhyay,
  M.~Mardani, T.~Kurth, D.~Hall, Z.~Li, K.~Azizzadenesheli, P.~Hassanzadeh,
  K.~Kashinath, A.~Anandkumar,
  \href{http://arxiv.org/abs/2202.11214}{Fourcastnet: A global data-driven
  high-resolution weather model using adaptive fourier neural operators},
  arXiv.2202.11214 (2 2022).
\newline\urlprefix\url{http://arxiv.org/abs/2202.11214}

\bibitem{Leith1972}
C.~E. Leith, R.~H. Kraichnan, Predictability of turbulent flows, J. Atmos. Sci.
  29 (1972) 1041--1058.

\bibitem{Lorenz}
E.~N. Lorenz, The essence of chaos (1993).

\bibitem{Vlachas2018}
P.~R. Vlachas, W.~Byeon, Z.~Y. Wan, T.~P. Sapsis, P.~Koumoutsakos, Data-driven
  forecasting of high-dimensional chaotic systems with long short-term memory
  networks, Proceedings of the Royal Society A: Mathematical, Physical and
  Engineering Sciences 474 (5 2018).
\newblock \href {https://doi.org/10.1098/rspa.2017.0844}
  {\path{doi:10.1098/rspa.2017.0844}}.

\bibitem{Li2020}
Z.~Li, N.~Kovachki, K.~Azizzadenesheli, B.~Liu, K.~Bhattacharya, A.~Stuart,
  A.~Anandkumar, \href{http://arxiv.org/abs/2010.08895}{Fourier neural operator
  for parametric partial differential equations}, arXiv.2010.08895 (10 2020).
\newline\urlprefix\url{http://arxiv.org/abs/2010.08895}

\bibitem{Vlachas2022}
P.~R. Vlachas, G.~Arampatzis, C.~Uhler, P.~Koumoutsakos, Multiscale simulations
  of complex systems by learning their effective dynamics, Nature Machine
  Intelligence 4 (2022) 359--366.
\newblock \href {https://doi.org/10.1038/s42256-022-00464-w}
  {\path{doi:10.1038/s42256-022-00464-w}}.

\bibitem{Chattopadhyay2020}
A.~Chattopadhyay, P.~Hassanzadeh, D.~Subramanian, Data-driven predictions of a
  multiscale lorenz 96 chaotic system using machine-learning methods: Reservoir
  computing, artificial neural network, and long short-term memory network,
  Nonlinear Processes in Geophysics 27 (2020) 373--389.
\newblock \href {https://doi.org/10.5194/npg-27-373-2020}
  {\path{doi:10.5194/npg-27-373-2020}}.

\bibitem{Vlachas2020}
P.~R. Vlachas, J.~Pathak, B.~R. Hunt, S.~T.~P., M.~Girvan, E.~Ott,
  P.~Koumoutsakos,
  \href{https://www.elsevier.com/open-access/userlicense/1.0/}{Backpropagation
  algorithms and reservoir computing in recurrent neural networks for the
  forecasting of complex spatiotemporal dynamics}, Neural Networks 126 (2020)
  191--217.
\newline\urlprefix\url{https://www.elsevier.com/open-access/userlicense/1.0/}

\bibitem{Pathak2018}
J.~Pathak, B.~Hunt, M.~Girvan, Z.~Lu, E.~Ott, Model-free prediction of large
  spatiotemporally chaotic systems from data: A reservoir computing approach,
  Physical Review Letters 120 (1 2018).
\newblock \href {https://doi.org/10.1103/PhysRevLett.120.024102}
  {\path{doi:10.1103/PhysRevLett.120.024102}}.

\bibitem{Yakhot1981}
V.~Yakhot, Large-scale properties of unstable systems governed by the Kuramoto-Sivashinksi equation,
  Physical Review A 24(1) (1981) 642--643.

\bibitem{Cross1993}
M.~C. Cross, P.~C. Hohenberg, Pattern formation outside of equilibrium, Reviews
  of Modern Physics 65 (1993) 851--1112.
\newblock \href {https://doi.org/10.1103/RevModPhys.65.851}
  {\path{doi:10.1103/RevModPhys.65.851}}.

\bibitem{Jaeger2004}
H.~Jaeger, H.~Haas, Harnessing nonlinearity: Predicting chaotic systems and
  saving energy in wireless communication, Science 304 (2004) 78--81.

\bibitem{Lukosevicius2012}
M.~Lukoševičius, A practical guide to applying echo state networks (2012).

\bibitem{Bratanov2013}
V.~Bratanov, F.~Jenko, D.~R. Hatch, M.~Wilczek, Nonuniversal power-law spectra
  in turbulent systems, Phys. Rev. Lett. 111 (2013) 075001.

\bibitem{Manneville1985}
P.~Manneville, Liapounov exponents for the kuramoto-sivashinsky model (1985).

\bibitem{Kassam2005}
A.~K. Kassam, L.~N. Trefethen, Fourth-order time-stepping for stiff pdes, SIAM
  Journal on Scientific Computing 26 (2005) 1214--1233.
\newblock \href {https://doi.org/10.1137/S1064827502410633}
  {\path{doi:10.1137/S1064827502410633}}.

\bibitem{Boffetta2002}
G.~Boffetta, M.~Cencini, M.~Falcioni, A.~Vulpiani, Predictability: a way to
  characterize complexity, Physics Reports 356 (2002) 367--474.

\bibitem{Majda2018}
A.~J. Majda, N.~Chen, Model error, information barriers, state estimation and
  prediction in complex multiscale systems, Entropy 20 (8 2018).
\newblock \href {https://doi.org/10.3390/e20090644}
  {\path{doi:10.3390/e20090644}}.

\bibitem{DelSole2004}
T.~DelSole, Predictability and information theory. part i: Measures of
  predictability, J. Atmos. Sci. 61 (2004) 2425--2440.

\bibitem{Goodfellow2014}
I.~J. Goodfellow, J.~Pouget-Abadie, M.~Mirza, B.~Xu, D.~Warde-Farley, S.~Ozair,
  A.~Courville, Y.~Bengio,
  \href{http://www.github.com/goodfeli/adversarial}{Generative adversarial
  nets}, Curran Associates, Inc., 2014.
\newline\urlprefix\url{http://www.github.com/goodfeli/adversarial}

\bibitem{Kim2021}
H.~Kim, J.~Kim, S.~Won, C.~Lee, \href{http://arxiv.org/abs/2007.15324
  http://dx.doi.org/10.1017/jfm.2020.1028}{Unsupervised deep learning for
  super-resolution reconstruction of turbulence}, J. Fluid Mech. 910 (2021)
  A29.
\newblock \href {https://doi.org/10.1017/jfm.2020.1028}
  {\path{doi:10.1017/jfm.2020.1028}}.
\newline\urlprefix\url{http://arxiv.org/abs/2007.15324
  http://dx.doi.org/10.1017/jfm.2020.1028}

\bibitem{Nishiura2001}
Y.~Nishiura, D.~Ueyama, Spatio-temporal chaos for the gray-scott model, Physica
  D 150 (2001) 137--162.

\bibitem{Driscoll2014}
T.~A. Driscoll, N.~Hale, L.~N. Trefethen, Chebfun Guide, Pafnuty Publications,
  2014.

\bibitem{Gauthier2021}
D.~J. Gauthier, E.~Bollt, A.~Griffith, W.~A. Barbosa, Next generation reservoir
  computing, Nature Communications 12 (12 2021).
\newblock \href {https://doi.org/10.1038/s41467-021-25801-2}
  {\path{doi:10.1038/s41467-021-25801-2}}.

\bibitem{Brunton2016}
S.~L. Brunton, J.~L. Proctor, J.~N. Kutz, W.~Bialek, Discovering governing
  equations from data by sparse identification of nonlinear dynamical systems,
  Proceedings of the National Academy of Sciences of the United States of
  America 113 (2016) 3932--3937.
\newblock \href {https://doi.org/10.1073/pnas.1517384113}
  {\path{doi:10.1073/pnas.1517384113}}.

\bibitem{Buhmann2000}
M.~D. Buhmann, Radial basis functions, Acta Numerica 9 (2000) 1--38.

\bibitem{Girosi1990}
F.~Girosi, T.~Poggio, Networks and the best approximation property, Biol.
  Cybern. 63 (1990) 169--176.

\bibitem{Zemouri2003}
R.~Zemouri, D.~Racoceanu, N.~Zerhouni, Recurrent radial basis function network
  for time-series prediction, Engineering Applications of Artificial
  Intelligence 16 (2003) 453--463.
\newblock \href {https://doi.org/10.1016/S0952-1976(03)00063-0}
  {\path{doi:10.1016/S0952-1976(03)00063-0}}.

\bibitem{Giloni2002}
A.~Giloni, M.~Padberg, Alternative methods of linear regression, Mathematical
  and Computer Modelling 35 (2002) 361--374.

\bibitem{Whitaker2009}
J.~S. Whitaker, G.~P. Compo, J.~N. Thépaut, A comparison of variational and
  ensemble-based data assimilation systems for reanalysis of sparse
  observations, Monthly Weather Review 137 (2009) 1991--1999.
\newblock \href {https://doi.org/10.1175/2008MWR2781.1}
  {\path{doi:10.1175/2008MWR2781.1}}.

\bibitem{Brajard2020}
J.~Brajard, A.~Carrassi, M.~Bocquet, L.~Bertino, Combining data assimilation
  and machine learning to emulate a dynamical model from sparse and noisy
  observations: A case study with the lorenz 96 model, Journal of Computational
  Science 44 (7 2020).
\newblock \href {https://doi.org/10.1016/j.jocs.2020.101171}
  {\path{doi:10.1016/j.jocs.2020.101171}}.

\bibitem{Takens1981}
F.~Takens, Detecting strange attractors in turbulence (1981).

\bibitem{Bakarji2022}
J.~Bakarji, K.~Champion, J.~N. Kutz, S.~L. Brunton,
  \href{http://arxiv.org/abs/2201.05136}{Discovering governing equations from
  partial measurements with deep delay autoencoders}, arXiv:2201.05136 (1
  2022).
\newline\urlprefix\url{http://arxiv.org/abs/2201.05136}

\bibitem{Raissi2020}
M.~Raissi, A.~Yazdani, G.~E. Karniadakis,
  \href{http://science.sciencemag.org/}{Hidden fluid mechanics: Learning
  velocity and pressure fields from flow visualizations}, Science 367 (2020)
  1026--1030.
\newline\urlprefix\url{http://science.sciencemag.org/}

\end{thebibliography}

\appendix

\section{Details of the reservoir networks}
\label{A1}

In an echo-state network (ESN), $W_{in} \in \mathbb{R}^{d_r\times d_u}$ connects the input vector $\mathbf{u}(t) \in \mathbb{R}^{d_u}$ to the reservoir nodes. Following \citep{Pathak2018}, each reservoir node is connected to a single element from the input vector, while each element of the input vector is connected to $d_r/d_u$ reservoir nodes. Connecting each reservoir node by a single input element ensures that uncorrelated input elements, such as from two spatially uncorrelated locations, do not get mixed in the reservoir. The non-zero elements of $W_{in}$ are randomly chosen from a normal distribution in [-1, 1]. These elements are then scaled by $\sigma$, which can be a scalar or a vector. In most of the networks used in the present study, $\sigma$ is a scalar. In KS$G_{rbf}$, GS$P^n$, GS$G_{rbf}$ and GS$2G_{rbf}$ networks, $\sigma \in \mathbb{R}^{L+1}$ (where $L$ is the number of nonlinear projectors) such that input from each nonlinear projector is scaled independently.

The adjacency matrix $A \in \mathbb{R}^{d_r\times d_r}$ determines the internal connections between the reservoir nodes. The elements of $A$ are also randomly chosen from a normal distribution in [-1, 1], they are then rescaled such that the largest eigenvalue of $A$ (also known as the spectral radius) is equal to $\rho$. It is sometimes recommended to make $A$ sparse by randomly fixing most of its elements to be equal to 0. However, we do not find any advantage in doing that in our case.

The elements of $W_{out} \in \mathbb{R}^{d_v\times d_r}$ are optimized to minimize the squared-error between the output $\mathbf{v}(t)$ and the target signal $\widehat{\mathbf{v}}(t)$. The optimized elements are obtained by performing the linear regression with Tikhonov regularization:
\begin{equation}
W_{out} = \widehat{\mathbf{V}}\mathbf{R}^T\left(\mathbf{R}\mathbf{R}^T + \beta\mathbf{I}\right)^{-1},
\end{equation}
where $\widehat{\mathbf{V}} = \left[\widehat{\mathbf{v}}(-T+N_{tran}\Delta t),\widehat{\mathbf{v}}(-T+N_{tran}\Delta t+\Delta t),...,\widehat{\mathbf{v}}(-\Delta t)\right]$, \\
$\mathbf{R} = \left[\mathbf{r}(-T+N_{tran}\Delta t), \mathbf{r}(-T+N_{tran}\Delta t+\Delta t),...,\mathbf{r}(-\Delta t)\right]$,
$\beta = 1e^{-5}$ is the regularization parameter and $\mathbf{I}$ is an identity matrix. The vectors $\widehat{\mathbf{v}}(-T)$, $\widehat{\mathbf{v}}(-T+\Delta t)$, ..., $\widehat{\mathbf{v}}(-\Delta t)$ and $\mathbf{r}(-T)$, $\mathbf{r}(-T+\Delta t)$, ..., $\mathbf{r}(-\Delta t)$ are the target signal and reservoir states during the training period, i.e. from $t = -T$ to 0, $N_{tran} = 50$ is the number of transient steps discarded during the training and $\Delta t$ is the discrete time step of the network.

\section{Calculations of the mean prediction error}
\label{A2}
There are two sources of randomness in the calculations of the prediction error (${\mathcal{E}(t)}$): (i) the initial condition of the system and (ii) the random seed used when drawing the elements of $W_{in}$ and $A$ matrices. Once the network is trained using data from $t=-T$ to 0, we calculate the predictions by reinitializing the network for 200 initial conditions. Given that ESNs satisfy the fading memory property, we only need a finite sequence of past inputs, say from $t_r - N_{ReInit}\Delta t$ to $t_r - \Delta t$, to obtain the predictions starting from $t_r$. We find that a finite sequence of $N_{ReInit} = 10$ past inputs is sufficient to reinitialize the ESNs in our study. During this reinitialization, the activation state is set to zero, i.e. $\mathbf{r}(t_r-N_{ReInit}\Delta t) = 0$, and is then evolved using the inputs $\mathbf{u}(t - N_{ReInit}\Delta t)$, $\mathbf{u}(t_r - N_{ReInit}\Delta t + \Delta t)$, ..., $\mathbf{u}(t_r-\Delta t)$. Such calculations are repeated for 10 random seeds. The mean prediction error ($\overline{\mathcal{E}(t)}$) is obtained by averaging the prediction error (${\mathcal{E}(t)}$) over 200 initial conditions and 10 random seeds, i.e. for 2000 calculations.

The mean prediction error ($\overline{\mathcal{E}(t)}$) presented in the paper is approximately optimized over all the hyperparameters for each network. The main hyperparameters in a network of given size $d_r$ and the number of parallel units $N_g$ are: (i) the interaction length $N_i$, (ii) the spectral radius $\rho$ of $A$ and (iii) the wieght $\sigma$ of input elements. When nonlinear projector functions are used, there are additional hyperparameters associated with the projector functions.

\section{Different radial basis functions}
\label{A3}
In the main text, we only used Gaussian radial basis functions (RBFs) as nonlinear projector for which we fixed the center at $y_c(x) = 0$ to predict the Gray-Scott system for which only concentration of one species, $y_1(x,t)$, is observable. Figure 2 shows that using Gaussian RBF with center at $y_c = 1$ and multiquadric RBF are equally good options, thus showing the robustness of RBFs as nonlinear projectors. A multiquadric RBF is given as $\sqrt{1 + \gamma\left(y_1-y_c\right)^2}$, where $\gamma$ is the shape factor and $y_c$ is the center.

\begin{figure}
 \centerline{\includegraphics[width=0.85\textwidth, trim =1.6cm 5.85cm 4.95cm 1.3cm, clip]{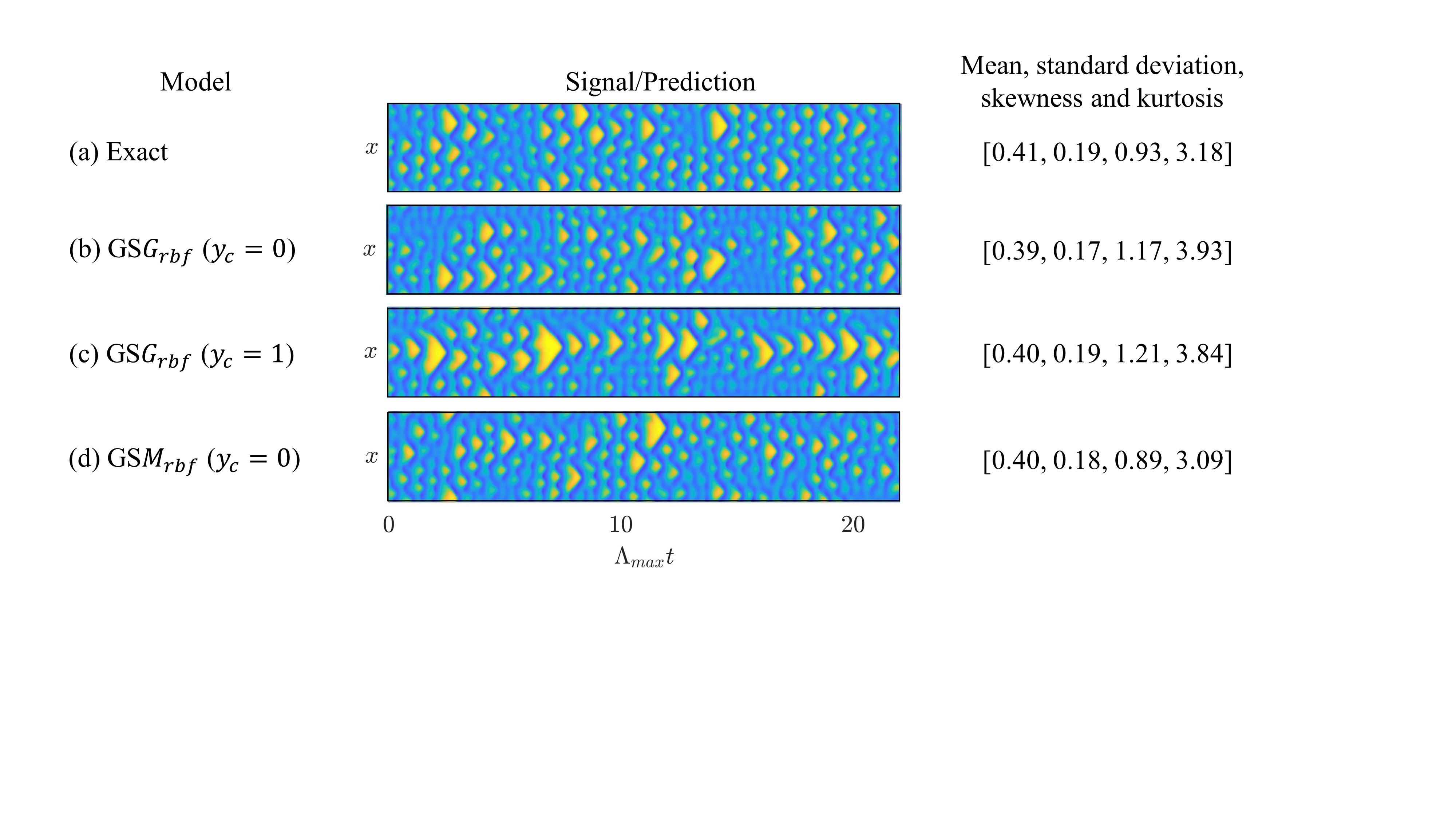}}
  \caption{(a) The observed signal $y_1$ of the Gray-Scott system and (b-d) its predictions from different RC-RNNs. The networks GS$G_{rbf}$ ($y_c = 0$) uses $\exp(-2y_1^2)$,  GS$G_{rbf}$ ($y_c = 1$) uses $\exp(-2(y_1-1)^2)$ and GS$M_{rbf}$ ($y_c = 0$) uses $\sqrt{1+3y_1^2}$ as nonlinear projectors.}
\label{Prbf}
\end{figure}





\end{document}